\documentclass[a4paper,12pt]{article}
\usepackage{graphicx,epsfig,color}
\usepackage{epstopdf}
\usepackage{indentfirst}
\usepackage{hyperref}
\usepackage{amsmath}
\usepackage[cp1251]{inputenc}
\usepackage[english]{babel}

\begin{document}
\sloppy

\inputencoding{cp1251}
\textwidth=160mm
\textheight=250mm
\voffset=-2cm
\linespread{0.8}

\title{Gas of thin tubes of massless scalar field as a possible model of unified dark fluid}

\author{Alexander Lelyakov$^{1\,*}$, Stepan Lelyakov$^{1}$ \\
$^{1}$Department of Theoretical Physics, Institute of Physics and\\
Technology, V.I.~Vernadsky Crimean Federal University,\\
 Simferopol 295007, Russia.}

\date{}
\maketitle
$^{*}$Corresponding author. E-mail: lelyakov\_a\_p@mail.ru

\date{}
\maketitle
\begin{description}
\item []
\end{description}

In this work, we investigated the influence of constants that characterize the motion of thin tubes of a massless scalar field (TToMSF) in a gas on the global properties of such a gas, in particular, on its ability to accelerated expansion. The peculiarity of the TToMSF gas is related to the fact that the parameter of its equation of state $w$ depends on the degree of rarefaction of the gas and can take values in a very wide range of values $-1< w <1$. In this case, fluctuations in the density of the TToMSF gas, caused by its gravitational interaction with non-uniformly distributed baryonic matter, should lead to the implementation of local equations of state with different values and different laws of change of the parameter $w$. The possibility of implementing local equations of state can alleviate the tensions associated with the observed non-uniformity of the distribution of matter and the growth of structures in the Universe.

\section{Introduction}\label{sec1}

The hypothesis of the presence of dark matter in our Universe arose as a result of the analysis of the rotation curves of observed spiral galaxies. Thus, it was noticed that the rotation curves of spiral galaxies do not agree with the observed density of baryonic matter, which follows from the function of their luminosity. At the same time, the anomalously high rotation speeds of the outer regions of galaxies could be explained by the contribution of some matter invisible to observers, which was later called dark matter \cite{ref-205}.

The understanding that the Universe is currently expanding at an accelerating rate came at the end of the 20th century, after two independent groups of astrophysicists measured the apparent magnitudes of type Ia supernovae (SNe Ia) at high redshift refs.~\cite{ref-206}-\cite{ref-2082}. The energy that is the source of the late accelerated expansion of the Universe is called dark energy (DE).

According to the Planck data release (PR4), approximately $95\%$ of the matter in the Universe is in the dark sector, i.e., it belongs to dark matter and dark energy, and only $\approx 4.86\%$ is visible baryonic matter  refs.~\cite{ref-200,ref-204}.

The $\Lambda$CDM and $\Lambda$WDM cosmological models, which assume the existence of cold (CDM) or warm (WDM) dark matter particles, respectively, are currently considered the standard models of cosmology (SMoC). Dark energy in these models is associated with the vacuum energy and is described by the cosmological constant $\Lambda$ in Einstein's equations refs.~\cite{ref-239}-\cite{ref-241}. The completeness of SMoC allows us to quantitatively evaluate and model the processes observed in the Universe, both at different scales and at different stages of its formation.
Comparative analysis of model results and obtained observational data, which span scales from star clusters to the cosmological horizon, revealed a number of fundamental tensions in SMoC. Below we present the series of SMoC tensions as was done in the paper by Kroupa, Gjergo et al. (see ref~\cite{ref-24101}):

($\bf T1$) Problem of galaxy shape. Galaxies in SMoC primarily grow through mergers, so most galaxies formed in SMoC simulations are bloated, bulge-dominated, or spheroidal  refs.~\cite{ref-24101,ref-2410}. In contrast, the real Universe is dominated by very thin disk galaxies, a significant fraction ($\approx30\%$) of which are devoid of bulges refs.~\cite{ref-2411}-\cite{ref-2414};

($\bf T2$) Lack of observational evidence for dynamical dissipation by extended dark matter halos. Due to existing strict experimental constraints, dark matter cannot be represented by particles that belong to the standard model of particle physics (SMoPP), but dark matter particles can interact with SMoPP particles gravitationally. It must be acknowledged that at the present time there is no experimental confirmation of the existence of dark matter particles \cite{ref-2415}. However, if dark matter particles exist, then, in accordance with the SMoC prediction, they will form massive and extended dark matter halos around the baryonic components of galaxies refs.~\cite{ref-24101}-\cite{ref-2417}. In this case, for local groups of galaxies, massive and extended dark matter halos, touching each other, should lead to dynamic Chandrasekhar friction (dissipation), as a result of which galaxies should slow down their relative speed, fall on each other and, eventually, merge into a single object  refs.~\cite{ref-24101,ref-2418}. Observations show that about 90 percent of all galaxies have thin, extended ancient disks of star formation, and these structures are not capable of sustaining repeated mergers that could be driven by dark matter \cite{ref-24101};

($\bf T3$) The problem of satellites. According to SMoC, galaxies in the Local Group should be distributed in a spheroidal configuration. Instead, observations show that galaxies in the Local Group are mostly found in thin planes. For example, the bulk of satellite galaxies for the Milky Way \cite{ref-301}-\cite{ref-303}, Andromeda \cite{ref-304,ref-305}, Centaurus A \cite{ref-306,ref-307}, M81 \cite{ref-308}, and others \cite{ref-309,ref-310} are located in huge rotating flat structures that resemble planetary systems, rather than forming spheroidal distributions as predicted by SMoC  refs.~\cite{ref-311,ref-312};

($\bf T4$) Hubble tension. It is well known that the value of the Hubble constant measured using SNe Ia differs significantly from the value that follows from the observed CMB. Namely, the measurement of SNe Ia gives the local value of the Hubble constant, $H_{0}^{local} = 73.04\pm 1.04 km s^{-1} Mpc^{-1}$ \cite{ref-313} ($H_{0}^{local} = 74.6\pm 3.0 km s^{-1} Mpc^{-1}$ \cite{ref-314}, $H_{0}^{local} = 76.9_{-4.8}^{+8.2} km s^{-1} Mpc^{-1}$ \cite{ref-315}). At the same time, the global expansion rate, which is measured by the CMB, is $H_{0}^{global} = 67.4\pm 0.5 km s^{-1} Mpc^{-1}$ \cite{ref-316}). The presence of a significant difference between the values of $H_{0}^{local}$ and $H_{0}^{global}$ is often associated with the need for alternative physical theories for dark energy, although the tension $\bf T4$ can have a natural explanation within the SMoC framework. In SMoC, the difference between the values of $H_{0}^{local}$ and $H_{0}^{global}$ can be explained by a region of reduced matter density at a scale of about $0.6\, Gpc$ (the Keenan-Barger-Cowie (KBC) void), which contains our Local Group of galaxies refs.~\cite{ref-317,ref-318};

($\bf T5$) Large-scale inhomogeneities of matter. The observable Universe is much more inhomogeneous on scales up to a few $Gpc$ than allowed by SMoC refs.~\cite{ref-315}-\cite{ref-322}.  The homogeneity scale of the SMoC universe is estimated to be about $100\, Mpc$ \cite{ref-318,ref-319}. At the same time, evidence for large-scale ($> 0.5\,Gpc$) inhomogeneities in the distribution of matter comes from observations of the motion of galaxy clusters \cite{ref-320}, the distribution of quasars and active galactic nuclei \cite{ref-321}, and the distribution of gamma-ray bursts \cite{ref-322}. The studies conducted have found that the putative superclusters are significantly larger in volume than allowed by SMoC. At the same time, the size of regions with reduced matter density, for example, KBC \cite{ref-323}-\cite{ref-326}, or Ho'oleilana \cite{ref-315}, are also in tension with SMoC;

($\bf T6$) Growth of structures in the Universe. Observations made with the James Webb Space Telescope (JWST) show that structures in the Universe are growing faster than SMoC allows. For example, the density of galaxies found at redshifts $z>14$ is approximately $100$ times higher than the SMoC predictions refs.~\cite{ref-326',ref-326''}. Galaxies with masses in the range $10^{9}-10^{11}\, M\odot$, discovered at redshifts $z> 10$ \cite{ref-327,ref-328}, the galaxy RUBIES-UDS-QG-z7 with a mass of about $10^{10}\, M\odot$ ($z=7.29$) \cite{ref-329}, the double cluster El Gordo with a mass of about $2\times10^{15}\, M\odot$ ($z=0.87$) \cite{ref-330,ref-331}, all these are examples confirming that real galaxies form and evolve much faster than predicted by SMoC refs.~\cite{ref-329,ref-24101};

($\bf T7$) The fine-tuning problem, or the problem of the cosmological constant $\Lambda$. The current understanding of the constant $\Lambda$ as the vacuum energy density leads to a difference of approximately $120$ orders of magnitude between the observed DE energy density and the corresponding prediction of quantum field theory (QFT) refs.~\cite{ref-250}-\cite{ref-335}. In QFT, the vacuum energy density is associated with the possibility of generating particle-antiparticle pairs. This understanding of vacuum energy is the basis for explaining the observed Casimir and Lamb effects refs.~\cite{ref-336,ref-337}. Theoretically, using the one-loop correction in QFT, the value of the cosmological constant $\Lambda$ can be made arbitrarily small. But, at the same time, the counterterms introduced with such a procedure are not protected from loop corrections of higher orders, which ultimately leads to an infinite tree of finely tuned counterterms, and in QFT there is no correct procedure that allows one to stop such branching.

A more complete and detailed analysis of the observational data that come into tension with SMoC can be found in the reviews \cite{ref-24101,ref-332}.

Attempts to alleviate the tensions in SMoC associated with dark energy have led to a large number of models with variable expansion dynamics of the Universe. Among such models we can name: models with parameterization \cite{ref-2631,ref-2632}, viscous models \cite{ref-2633}-\cite{ref-2635}, the emergent dark energy model \cite{ref-2636,ref-2637}, modified gravity models \cite{ref-2638}-\cite{ref-26314}, dynamic scalar field models \cite{ref-276}-\cite{ref-279}. For example, for models of the dynamic scalar field of dark energy, the parameter of the equation of state $w=p/\rho$, where $p$ and $\rho$ are, respectively, the pressure and energy density, is a variable quantity that depends on time.

The concept of a thin tube of massless scalar field (TToMSF) is inextricably linked with such concepts in cosmology as the classical cosmic string and the null string. Classical cosmic strings are ``one-dimensional'' topological defects that, according to existing ideas, could have arisen during phase transitions in the early Universe refs.~\cite{ref-1}-\cite{ref-5}. At the same time, null strings realize the limit in which the speed of a classical cosmic string reaches the speed of light, while the rest mass of the cosmic string tends to zero, and its energy remains a finite value refs.~\cite{ref-274,ref-275}. For this reason, the second name for null strings that appears in the literature is massless cosmic string.

When describing the motion of a null string in an external gravitational field, an approximation is usually used in which the null string is replaced by a one-dimensional model. However, the one-dimensional model for the null string is not always physically justified since it corresponds to the study on a very large spatial scale, i.e., it corresponds to the scale of observation at which the gravitational influence of the null string is vanishingly small. In this regard, when studying the gravitational influence of an individual null string in a gas on its surroundings, it is physically justified to apply not a one-dimensional null string model, but the TToMSF model refs.~\cite{ref-7}-\cite{ref-101}.

The equation of state parameter $w$ for the TToMSF gas is a variable value and depends on the degree of rarefaction of such a gas, i.e., it changes over time. Thus, in the case of a super-compressed gas $w\rightarrow1$, and for a highly rarefied $w\rightarrow-1$ refs.~\cite{ref-270,ref-2701}. Based on this feature, TToMSF gas can be classified as a model of a dynamic scalar field of dark energy.

The advantages of the TToMSF model over other models are:

($\bf A1$) Time of formation. In accordance with the historically established concept of the formation of classical cosmic strings and null strings, which was noted above, the TToMSF gas could have formed during the period of baryogenesis ($t\approx10^{-35}\,s$ ref.~\cite{ref-1}). That is, such a gas could have formed already at the earliest stages of the evolution of the Universe and, thus, could have influenced all processes and phenomena in the Universe.

($\bf A2$) Possibility of implementing local equations of state. As noted above, the equation of state parameter $w$ for the TToMSF gas depends on the degree of rarefaction of this gas and can take values in a very wide range $-1< w <1$ refs.~\cite{ref-270,ref-2701}. We can also assume two possible connections of TToMSF gas with baryonic matter, namely, ``weak'' and ``strong''. With ``weak'' coupling, TToMSF gas interacts with baryonic matter only gravitationally. That is, for a ``weak'' coupling, the TToMSF gas is gravitationally entrained by the baryonic matter and, thus, the density of the TToMSF gas will repeat the inhomogeneities in the distribution density of the baryonic matter. Since the density of the gas is directly related to the rarefaction, then the denser and less dense regions of the distribution of baryonic matter will correspond to the equations of state of the gas TToMSF with different values of the parameter $w$. For example, in the epoch of baryon acoustic oscillations (BAO), the implementation of local equations of state will lead to the fact that the rarefied regions will be larger and the density of matter in them will be lower than predicted by SMoC. Conversely, denser regions will be thinner and the matter concentration in them will be greater than predicted by SMoC, or any other alternative cosmological model with a global equation of state. Further evolution of inhomogeneities with different local equations of state will enable faster growth of the structure for initially denser regions, and vice versa, the growth of structures in initially less dense regions will be suppressed.

Thus, the possibility of implementing local equations of state can soften the tensions $\bf T5$ and $\bf T6$, which are given above. We will discuss the concept and possibility of realizing a ``strong'' connection between the TToMSF gas and baryonic matter in the ``Discussion'' chapter.

In this work, we investigated the influence of constants that characterize the gravitational field and the motion of TToMSF in gas on the global properties of such gas, in particular, on its ability to accelerated expansion. We investigated a gas whose elements are closed TToMSFs whose spatial shape satisfies axial symmetry and which increase in size (``radius'') with increasing time, while maintaining their initial shape unchanged. In section~\ref{sec2}, we presented a distribution function that models a solitary TToMSF, and presented metric functions that describe the gravitational field of a TToMSF with a given type of motion and spatial symmetry. In section~\ref{sec3}, we found conditions under which the test null string model is physically justified. In section~\ref{sec4}, we found and investigated solutions to the equations of motion of a test null string in the gravitational field TToMSF (another name applied to TToMSF is source string). In section~\ref{sec5}, we presented graphs that characterize the motion of a test null string in the gravitational field TToMSF. In section~\ref{sec6} we discussed the influence of the gravitational field of a TToMSF on the motion of neighboring TToMSFs in a gas, and also discussed the influence of the constants that characterize the gravitational field and the motion of thin tubes of a massless scalar field in a gas on the equation of state of such a gas.

\section{Gravitational field of a source string}\label{sec2}

In this work, we investigated the gravitational influence of a null string (TToMSF), which, over a some time interval, radially increases its size and, at the same time, maintains its initial shape unchanged.

Let a cylindrical coordinate system $x^{0}=t$, $x^{1}=r$, $x^{2}=\theta$, $x^{3}=z$ be defined in space. In this coordinate system, the functions that define the world surface (trajectory of motion) of a one-dimensional closed null-string model, which we will further consider to be the source of the gravitational field, have the form:
\begin{equation}
\label{eq:1}
t = \tilde{\tau},  \quad r_{source} = \tilde{\tau},  \quad \theta_{source} = \tilde{\sigma},  \quad z_{source} = q(\theta),
\end{equation}
where: $\tilde{\tau}$ and $\tilde{\sigma}$ are the parameters on the world surface of the null string, $\tilde{\tau}\in\left[0, \tilde{\tau}_{0}\right]$, $\tilde{\sigma}\in\left[0; 2\pi\right]$, $\tilde{\tau}_{0}$ is a constant, the function $q(\theta)$ defines the initial shape of the null string, and satisfies the following constraints:

$\bullet$ closedness condition for the null string
$$
q(\theta)=q(\theta+2\pi);
$$

$\bullet$ the invariance condition under the inversion $\theta$ on $-\theta$, i.e.,
$$
q(\theta)=q(-\theta).
$$

\begin{figure}[ht]
\center{\includegraphics[width=8cm]{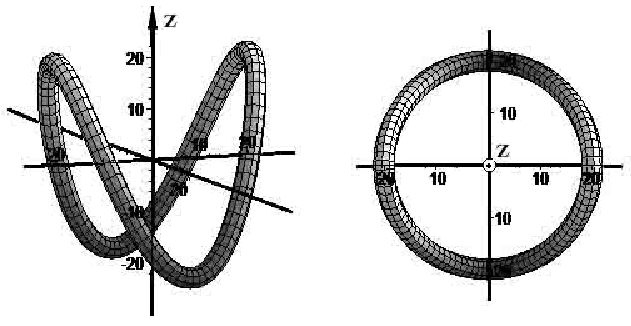}}
\hfill
\caption{\footnotesize The figure shows the configuration of a closed null string (TToMSF), which corresponds to the function $q(\theta)=20\cos(2\theta)$.}
\label{f1}
\end{figure}

As already mentioned, when studying the gravitational field of a solitary null string, it is convenient to move from the null string model as a one-dimensional object to the null string model in the form of TToMSF. Figure~\ref{f1} shows an example of a possible configuration of a closed TToMSF that satisfies the conditions of the problem and corresponds to the function $q(\theta)=20\cos(2\theta)$.

The metric describing the gravitational field TToMSF, which corresponds to the motion of its one-dimensional analogue along the trajectory (\ref{eq:1}), can be represented as (see refs.~\cite{ref-79,ref-10}):
\begin{equation}
\label{eq:2}
dS^{2}=e^{2\nu}\left((dt)^{2} - (dr)^{2}\right) - B(d\theta)^{2} - e^{2\mu}(dz)^{2},
\end{equation}
where
\begin{equation}
\label{eq:3}
e^{2\nu}=a_{0}|\lambda_{,\eta}|(\alpha(\eta)+\lambda(\eta)f(Q)),
\end{equation}
\begin{equation}
\label{eq:4}
B=b_{0}\lambda^{2}(\eta)(\alpha(\eta)+\lambda(\eta)f(Q))\frac{(q_{,\theta})^{2}\left(f_{,Q}\right)^{2}}{\left[1-\sqrt{1-\left(c_{0}f_{,Q}\right)^{2}}\right]},
\end{equation}
\begin{equation}
\label{eq:5}
e^{2\mu}=b_{0}\lambda^{2}(\eta)(\alpha(\eta)+\lambda(\eta)f(Q))\frac{\left(f_{,Q}\right)^{2}}{\left[1+\sqrt{1-\left(c_{0}f_{,Q}\right)^{2}}\right]}.
\end{equation}
In equalities (\ref{eq:3}) -- (\ref{eq:5}), the quantities $a_{0},b_{0},c_{0}$ are constants,
\begin{equation}
\label{eq:5'}
\eta=t-r,
\end{equation}
\begin{equation}
\label{eq:5''}
Q=Q(z, \theta)=z-q(\theta).
\end{equation}
Note that from eq. (\ref{eq:1}) it follows that for $r\rightarrow r_{source}$ the variable $\eta\rightarrow 0$, and for $z\rightarrow z_{source}$ the variable $Q\rightarrow 0$.

The functions $\lambda(\eta)$ and $\alpha(\eta)$ are related to each other by the equality
\begin{equation}
\label{eq:7}
\lambda(\eta)=(1-\alpha(\eta))/f_{0}, \quad f_{0}=const.
\end{equation}

The functions $\alpha(\eta)$, $\lambda(\eta)$, and $f(Q)$ define the distribution function of a scalar field:
\begin{equation}
\label{eq:6}
  \varphi(\eta, z, \theta)=-\ln\left(\left(\alpha(\eta)+\lambda(\eta)f(Q)\right)^{\gamma}\right),
\end{equation}
where $\gamma=\sqrt{3/2\chi}$, $\chi=8\pi G$ (in the system of units $c=1$, where $c$ is the speed of light), $G$ is the gravitational constant.

In the limit of compression of TToMSF into a one-dimensional object (null string), the following conditions must be met \cite{ref-10}:
\begin{equation}
\label{eq:8}
\left.(\varphi_{,\eta})^{2}\right\vert_{\eta\rightarrow 0,\,Q\rightarrow0}\rightarrow \infty, \quad  \left.\varphi_{,Q}\right\vert_{\eta\rightarrow 0,\,Q\rightarrow0} \rightarrow 0,\quad
\left.\varphi_{,\eta}\varphi_{,Q}\right\vert_{\eta\rightarrow 0,\,Q\rightarrow0} \rightarrow 0,
\end{equation}
at the same time, in the space that surrounds TToMSF, the following conditions must be met \cite{ref-10}:
\begin{equation}
\label{eq:8''}
  \varphi \rightarrow 0,\quad \varphi_{,\eta}\rightarrow 0, \quad \varphi_{,Q}\rightarrow 0.
\end{equation}

Below is an example of functions $\alpha(\eta)$ and $f(Q)$ that can satisfy conditions (\ref{eq:8}), (\ref{eq:8''}):
\begin{equation}
\label{eq:9}
\alpha(\eta)=\exp\left(-(\xi (\vert\eta\vert+\epsilon'/\xi))^{-2}\right),
\end{equation}
\begin{equation}
\label{eq:10}
f(Q)=f_{0}\exp\left(-\varsigma\left(1-\exp\left(-\left(\zeta (\vert Q\vert+\epsilon/\zeta)\right)^{-2}\right)\right)\right),
\end{equation}
where the constants $\xi$ and $\zeta$ determine the ``thickness'' of TToMSF, and the positive constants $\epsilon$, $\epsilon'$, and $\varsigma$ ensure that the conditions (\ref{eq:8}) are satisfied. Namely, as follows from (\ref{eq:9}), (\ref{eq:10}), when compressing a scalar field into a one-dimensional object (null string):
\begin{equation}
\label{eq:11}
\xi\rightarrow\infty, \quad \zeta\rightarrow\infty, \quad \epsilon\rightarrow0, \quad \epsilon'\rightarrow0, \quad \varsigma\rightarrow\infty.
\end{equation}

In the system of units $c=1$, $G=1$, where $c$ is the speed of light, the constants $f_{0}$, $\chi$, $\varsigma$, $\gamma$, $\epsilon$, $\epsilon'$, are dimensionless. The dimensions of the constants: $a_{0}$, $c_{0}$, are $m$. The dimensions of the constants: $\xi$, $\zeta$, are $m^{-1}$. The dimension of the constant $b_{0}$ is $m^{2}$.

\begin{figure}[ht]
\center{\includegraphics[width=5cm]{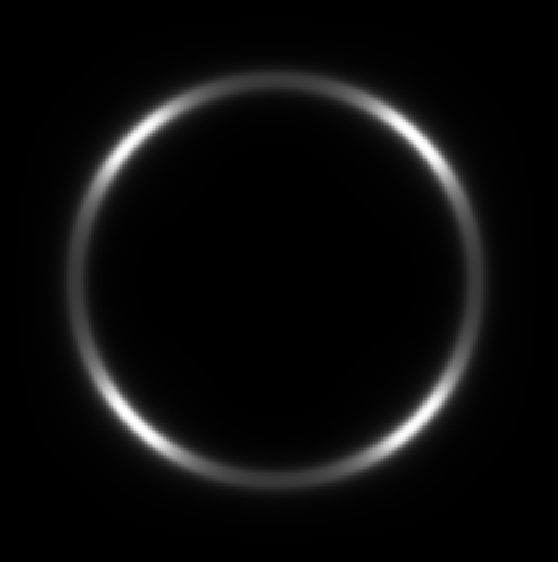}}
\hfill
\caption{\footnotesize The figure shows the distribution of the scalar field on the surface $z=0$, for values $\xi=\zeta=2$.}
\label{f4}
\end{figure}

\begin{figure}[ht]
\center{\includegraphics[width=5cm]{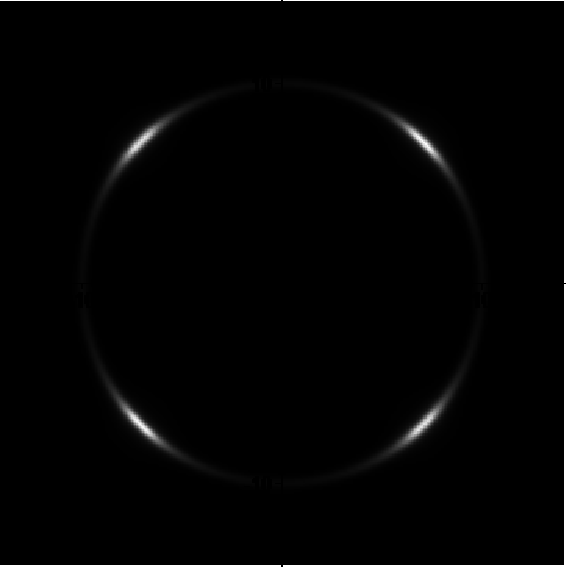}}
\hfill
\caption{\footnotesize The figure shows the distribution of the scalar field on the surface $z=0$, for values $\xi=\zeta=5$.}
\label{f6}
\end{figure}

In Figure~\ref{f4} and Figure~\ref{f6}, for different values of the constants $\xi$ and $\zeta$, and the functions $\alpha(\eta)$ and $f(Q)$, given by eqs.~(\ref{eq:9}), (\ref{eq:10}), the function $q(\theta)=\cos(2\theta)$, the distributions of the scalar field (\ref{eq:6}) on the hypersurface $z=0$, at time $t=10$ are presented. In these figures, the area where $\varphi\rightarrow 0$ is highlighted in black.

From the figures provided it is evident that with an increase in the value of the constants $\xi$ and $\zeta$, the region in which the value of the scalar field $\varphi$ is different from zero decreases, i.e., the ''thickness'' of TToMSF, in which the scalar field is concentrated, decreases.

\section{Model of test null string - conditions of applicability}\label{sec3}

In this paper we will investigate the motion of a test null string in the gravitational field of another null string (a thin tube of massless scalar field), which is the source of the gravitational field. In fact, the problem studies a system consisting of two closed null strings, while the gravitational contribution of one of the null strings, namely, the gravitational influence of the test null string, is neglected. It is important to find out the conditions under which such a situation can be realized, i.e., to find out the conditions under which the model of a test null string is physically justified.

The conditions under which the gravitational contribution of one of the strings can be neglected can be determined by studying the invariants of the quadratic form (\ref{eq:2}). As such an invariant, it is convenient to choose the expression for the scalar curvature: $K=g^{ij}g^{\mu\nu}R_{i\mu j\nu}$, where $R_{i\mu j\nu}$ is the Riemann-Christoffel tensor. For the quadratic form (\ref{eq:2}) and metric functions (\ref{eq:3}) -- (\ref{eq:5}), the expression for the scalar curvature $K$ has the form
\begin{equation}
\label{eq:105}
K=-\chi\left[\frac{(\varphi_{,z})^2}{e^{2\mu}}+\frac{(\varphi_{,\theta})^2}{B}\right]=
\frac{-3}{b_{0}(\alpha(\eta)+\lambda(\eta)f(Q))^{3}}.
\end{equation}

In the region that surrounds the null string $\varphi\rightarrow0$, so (see eq. (\ref{eq:6}))
\begin{equation}
\label{eq:105'}
(\alpha(\eta)+\lambda(\eta)f(Q))\rightarrow1.
\end{equation}
Inside a thin tube of the scalar field (for $\eta\rightarrow0$, $Q\rightarrow0$), applying eqs. (\ref{eq:7}), (\ref{eq:9}), (\ref{eq:10}), (\ref{eq:11}), we find
\begin{equation}
\label{eq:105''}
(\alpha(\eta)+\lambda(\eta)f(Q))\rightarrow \Phi\rightarrow0,
\end{equation}
where
\begin{equation}
\label{eq:105'''}
\Phi=\exp(-1/\epsilon'^2)+(1-\exp(-1/\epsilon'^2))\exp(-\varsigma\left(1-\exp(-1/\epsilon^2)\right)).
\end{equation}

Applying (\ref{eq:105'}), (\ref{eq:105''}) to (\ref{eq:105}), we obtain that in the region that surrounds the null string
\begin{equation}
\label{eq:106}
K_{out}\rightarrow-\frac{3}{b_{0}},
\end{equation}
and inside a thin tube of scalar field
\begin{equation}
\label{eq:107}
K_{within}\rightarrow-\frac{3}{b_{0}\Phi^{3}}.
\end{equation}

Using the results of ref.~\cite{ref-10}, one can show that for a null string (one-dimensional object) moving along the trajectory (\ref{eq:1}) the scalar curvature $K$ is equal to zero in the entire space ($K=0$). In other words, as the thickness of the scalar field tube decreases, the scalar curvature of the space $K$ must decrease (in absolute value). Moreover, with the limiting transition from a thin tube of a scalar field to a one-dimensional object (null string), the scalar curvature ($K$) should tend to zero in the entire space.  In this connection, in eqs. (\ref{eq:106}), (\ref{eq:107}), it is necessary to correctly determine the value of the constant $b_{0}$.

As shown in ref.~\cite{ref-10}, the constants $c_{0}$ and $b_{0}$ in the functions (\ref{eq:4}), (\ref{eq:5}) are related to each other by the equality
\begin{equation}
\label{eq:99}
b_{0}=b_{1}c_{0},
\end{equation}
where $b_{1}$ is a constant (one of the constants of integration of Einstein's equations). The dimension of the constant $b_{1}$ is $m$ (in the system of units $c=1$, $G=1$). At the same time (see eqs. (\ref{eq:4}), (\ref{eq:5})), the value of the constant $c_{0}$ must ensure that the conditions are met
\begin{equation}
\label{eq:99'}
0< \left(c_{0}f_{,Q}\right)^{2}\leq 1.
\end{equation}

For a function $f(Q)$ defined by eq.~(\ref{eq:10}), the maximum value of the function $f_{,Q}$ is achieved when
\begin{equation}
\label{eq:103'''}
\left(\zeta (Q+\epsilon/\zeta)\right)^{-2}=\frac{3}{2(\varsigma+1)}.
\end{equation}
For (\ref{eq:103'''}), the value of the function $f_{,Q}$ can be estimated as
\begin{equation}
\label{eq:104'}
\vert f_{,Q}\vert <2f_{0}\varsigma\zeta e^{-\varsigma}\exp\left(\varsigma e^{-3/(2(\varsigma+1))}\right)\left(\frac{3}{2(\varsigma+1)}\right)^{3/2}.
\end{equation}

Then, the value of the constant $c_{0}$, at which the constraints (\ref{eq:99'}) will be satisfied, can be fixed by the equality
\begin{equation}
\label{eq:104}
c_{0}=\left(2f_{0}\varsigma\zeta e^{-\varsigma}\exp\left(\varsigma e^{-3/(2(\varsigma+1))}\right)\left(\frac{3}{2(\varsigma+1)}\right)^{3/2}\right)^{-1}.
\end{equation}
Taking into account (\ref{eq:106}), (\ref{eq:107}), (\ref{eq:99}), and (\ref{eq:104}), we can say that the scalar curvature of the space $K$ is determined by constants that characterize the thickness of the thin tube of the scalar field.

For a massless object, the curvature of space cannot have singularities and the gravitational influence at an infinite distance from the null string must be absent, therefore the constant $b_{1}$, in eq. (\ref{eq:99}), taking into account (\ref{eq:104}), for example, can be fixed by the equality
\begin{equation}
\label{eq:108}
b_{1}=2\varrho f_{0}\varsigma\zeta^{2}\xi e^{2\varsigma}\exp\left(\varsigma e^{-3/(2(\varsigma+1))}\right)\left(\frac{3}{2(\varsigma+1)}\right)^{3/2},
\end{equation}
where $\varrho$ is a positive constant, and $\varrho>1$. The dimension of the constant $\varrho$ is $m^{4}$ (in the system of units $c=1$, $G=1$).

Applying (\ref{eq:104}), (\ref{eq:108}), for (\ref{eq:99}), we find an expression for the constant $b_{0}$, namely,
\begin{equation}
\label{eq:108'}
b_{0}=\varrho\zeta\xi e^{3\varsigma}.
\end{equation}

For (\ref{eq:108'}), eqs. (\ref{eq:106}), (\ref{eq:107}), respectively, take the form
\begin{equation}
\label{eq:109}
K_{out}\rightarrow-\frac{3e^{-3\varsigma}}{\varrho\zeta\xi}, \quad K_{within}\rightarrow-\frac{3e^{-3\varsigma}}{\varrho\zeta\xi\Phi^3}.
\end{equation}
Note that under the condition $1/\epsilon'^{2}>>\varsigma$, eq. (\ref{eq:105'''}), neglecting higher-order smallnesses, can be written as
\begin{equation}
\label{eq:110}
\Phi\approx e^{-\varsigma}.
\end{equation}
Taking into account (\ref{eq:110}), we can finally write
\begin{equation}
\label{eq:109'}
K_{out}\rightarrow-\frac{3e^{-3\varsigma}}{\varrho\zeta\xi}, \quad K_{within}\rightarrow-\frac{3}{\varrho\zeta\xi}.
\end{equation}

As was said earlier, the constants $\xi$ and $\zeta$ determine the ``thickness'' (degree of compression) of the tube inside which the scalar field is concentrated.
According to (\ref{eq:11}), (\ref{eq:109'}) at the limit transition from a thin tube of a scalar field to a one-dimensional object (null string) $K_{out}\rightarrow0$ and $K_{within}\rightarrow0$. At the same time (see eq. (\ref{eq:109'})), the condition will always be met
$$
K_{out}/K_{within}\approx e^{-3\varsigma}\rightarrow0.
$$

Let us note that the proposed fixation of the constants $c_{0}$ and $b_{1}$ is not the only possible one. It is possible to propose other expressions for the constants $c_{0}$, $b_{1}$, for which, in the limiting transition from a thin tube of a scalar field to a one-dimensional object (null string) $K_{out}\rightarrow0$ and $K_{within}\rightarrow0$.

Thus, the application of the test null string model will be physically justified if the values of the constants that characterize the test null string and the null string that is the source of the gravitational field (i.e., a thin tube of massless scalar field) are related by the following conditions
\begin{equation}
\label{eq:111}
(\zeta; \, \xi; \, \varsigma)_{test}>>(\zeta; \, \xi; \, \varsigma)_{source}, \quad (\epsilon; \, \epsilon')_{test}<<(\epsilon; \, \epsilon')_{source}.
\end{equation}
In other words, if the conditions (\ref{eq:111}) are met, the gravitational influence of the test null string on the null string, which is the source of the gravitational field, can be neglected. According to (\ref{eq:11}), (\ref{eq:111}) the ``thickness'' of the test null string must be less than the ``thickness'' of the null string that is the source of the gravitational field.

Below, as an example, for the values of the constants
\begin{equation}
\label{eq:111'}
\varrho=3, \quad \zeta=\varsigma=\xi=f_{0}=5,
\end{equation}
the values of the constants $c_{0}$, $b_{1}$, $b_{0}$, which are defined by the eqs. (\ref{eq:104}), (\ref{eq:108}), (\ref{eq:108'}), are
\begin{equation}
\label{eq:111''}
c_{0}\approx 0.0967, \quad b_{1}\approx 2\,535\,432\,006, \quad b_{0}\approx 245\,176\,275.
\end{equation}

\section{Motion of the test null string}\label{sec4}

The equations that describe the motion of a test null string in an external gravitational field with metric tensor $g_{ij}$ can be reduced to the form \cite{ref-6}:
\begin{equation}
\label{eq:13}
x^{\alpha}_{ ,\tau\tau}+\Gamma^{\alpha}_{pq}x^{p}_{,\tau}x^{q}_{,\tau}=0,
\end{equation}
\begin{equation}
\label{eq:14}
g_{\alpha\beta}x^{\alpha}_{,\tau}x^{\beta}_{,\tau}=0,
\end{equation}
\begin{equation}
\label{eq:14a}
g_{\alpha\beta}x^{\alpha}_{,\tau}x^{\beta}_{,\sigma}=0,
\end{equation}
where $\Gamma^{\alpha}_{pq}$ are the Christoffel symbols for the tensor $g_{ij}$, $\tau$ and $\sigma$ are the parameters on the world surface of the test null string, $x^{\alpha}_{,\tau}=\partial x^{\alpha}/\partial \tau$, $x^{\beta}_{,\sigma}=\partial x^{\beta}/\partial \sigma$, indices $\alpha$, $\beta$, $p$, $q$, $i$, $j$ take values 0,1,2,3, functions $x^{\alpha}(\tau, \sigma)$ determine the trajectory of the null string (world surface). At the same time, eq. (\ref{eq:14}) is the condition for the null string to be massless (i.e., the condition that all points of the null string, at each instant of time, move at the speed of light), and eq. (\ref{eq:14a}) ensures the orthogonality of the parameters $\tau$ and $\sigma$ on the world surface of the null string.

If the initial value of the polar angle $\theta$ for each point of the test closed null string does not change over time (i.e., does not depend on the time-like parameter $\tau$), then for the quadratic form (\ref{eq:2}) the equations of motion of the test null string (\ref{eq:13}) - (\ref{eq:14a}) can be reduced to the form:
\begin{equation}
\label{eq:22}
\theta=\theta(\sigma),
\end{equation}
\begin{equation}
\label{eq:15}
\eta_{,\tau\tau}+2\nu_{,\tau}\eta_{,\tau}=0,
\end{equation}
\begin{equation}
\label{eq:23}
s_{,\tau\tau}+2\nu_{,Q}Q_{,\tau}s_{,\tau}+2e^{2(\mu-\nu)}\mu_{,\eta}(z_{,\tau})^{2}=0,
\end{equation}
\begin{equation}
\label{eq:24}
z_{,\tau\tau}+e^{-2\mu}\left(e^{2\mu}\right)_{,\tau}z_{,\tau}+\frac{1}{2}e^{-2\mu}\left\{\left(e^{2\nu}\right)_{,Q}\eta_{,\tau}s_{,\tau}-
\left(e^{2\mu}\right)_{,Q}(z_{,\tau})^{2}\right\}=0,
\end{equation}
\begin{equation}
\label{eq:25}
\left(e^{2\nu}\right)_{,\theta}\eta_{,\tau}s_{,\tau}-\left(e^{2\mu}\right)_{,\theta}(z_{,\tau})^{2}=0,
\end{equation}
\begin{equation}
\label{eq:26}
e^{2\nu}\eta_{,\tau}s_{,\tau}-e^{2\mu}(z_{,\tau})^{2}=0,
\end{equation}
\begin{equation}
\label{eq:27}
e^{2\nu}\left(\eta_{,\tau}s_{,\sigma}+\eta_{,\sigma}s_{,\tau}\right)-2e^{2\mu}z_{,\tau}z_{,\sigma}=0,
\end{equation}
where
\begin{equation}
\label{eq:21}
s=t+r.
\end{equation}

In the region where the quadratic form (\ref{eq:2}) is defined, the functions $e^{2\mu}\neq0$ and $(e^{2\mu})_{,\theta}\neq0$. Then in this region, eqs. (\ref{eq:25}), (\ref{eq:26}), can be written as
\begin{equation}
\label{eq:28}
(z_{,\tau})^{2}=\frac{\left(e^{2\nu}\right)_{,\theta}\eta_{,\tau}s_{,\tau}}{\left(e^{2\mu}\right)_{,\theta}},
\end{equation}
\begin{equation}
\label{eq:29}
(z_{,\tau})^{2}=e^{2(\nu-\mu)}\eta_{,\tau}s_{,\tau}.
\end{equation}

It should be noted that for physically justified trajectories of the test null string, which are different from the trajectory of the source string (\ref{eq:1}), the conditions will be satisfied
\begin{equation}
\label{eq:30}
\eta_{,\tau}>0, \quad s_{,\tau}>0.
\end{equation}
The condition $\eta_{,\tau}<0$, or $s_{,\tau}<0$, corresponds to the case of radial expansion or, respectively, radial compression of the test null string at a speed greater than the speed of light, which has no physical justification. At the same time, the condition $\eta_{,\tau}=0$ corresponds to the trajectories of the test null strings, which are similar to the trajectory of the source string (\ref{eq:1}).

Equating the right-hand sides of eqs. (\ref{eq:28}), (\ref{eq:29}), and taking into account eq. (\ref{eq:30}), we find
\begin{equation}
\label{eq:31}
\left(\nu-\mu\right)_{,\theta}=0.
\end{equation}
For functions (\ref{eq:3}), (\ref{eq:5}), eq. (\ref{eq:31}) takes the form
\begin{equation}
\label{eq:32'}
\left(\ln{\left\{1-\sqrt{1-\left(c_{0}f_{,Q}\right)^{2}}\right\}}\right)_{,\theta}=0,
\end{equation}
or, given (\ref{eq:22}),
\begin{equation}
\label{eq:32}
\left(\ln{\left\{1-\sqrt{1-\left(c_{0}f_{,Q}\right)^{2}}\right\}}\right)_{,\sigma}=0.
\end{equation}
According to eq.~(\ref{eq:10}), the function $f_{,Q}$ depends on the variable $Q$, then the consequence of eq.~(\ref{eq:32}) is
\begin{equation}
\label{eq:33}
Q=Q(\tau).
\end{equation}
Since the form of the function $Q$ is determined by the eq. (\ref{eq:5''}), then to satisfy the eq. (\ref{eq:33}), taking into account (\ref{eq:22}), it is necessary to require
\begin{equation}
\label{eq:34}
z(\tau,\sigma)=Q(\tau)+q(\sigma).
\end{equation}

For (\ref{eq:30}), taking into account (\ref{eq:29}), (\ref{eq:33}), (\ref{eq:34}), the first integrals of eqs. (\ref{eq:15}), (\ref{eq:23}), (\ref{eq:24}) can be written as
\begin{equation}
\label{eq:42}
\eta_{,\tau}=P_{1}(\sigma)e^{-2\nu},
\end{equation}
\begin{equation}
\label{eq:47}
s_{,\tau}=a_{0}\frac{P_{3}(\sigma)}{P_{1}(\sigma)}\frac{\left|\lambda_{,\eta}\right|}{(\lambda(\eta))^{2}}\eta_{,\tau},
\end{equation}
\begin{equation}
\label{eq:46}
\frac{\left|f_{,Q}\right|\left|Q_{,\tau}\right|}{\left(1+\sqrt{1-\left(c_{0}f_{,Q}\right)^{2}}\right)^{1/2}}=
\frac{a_{0}c_{0}}{b_{0}}\frac{P_{2}(\sigma)}{P_{1}(\sigma)}\frac{\left|\lambda_{,\eta}\right|}{(\lambda(\eta))^{2}}\eta_{,\tau}.
\end{equation}
where $P_{1}(\sigma)$, $P_{2}(\sigma)$, $P_{3}(\sigma)$, are ``constants'' of integration (initial impulses of the points of the test null string), and, taking into account (\ref{eq:30}),
\begin{equation}
\label{eq:45}
P_{1}(\sigma)>0, \quad P_{2}(\sigma)>0, \quad P_{3}(\sigma)>0.
\end{equation}

According to (\ref{eq:10}), (\ref{eq:33}), the left side of the eq. (\ref{eq:46}) depends only on the variable $\tau$, in this case the right side of the eq. (\ref{eq:46}) can also depend only on the variable $\tau$, i.e., we can write
\begin{equation}
\label{eq:48}
\frac{P_{2}(\sigma)}{P_{1}(\sigma)}\frac{\left\vert\lambda_{,\eta}\right\vert}{(\lambda(\eta))^{2}}\eta_{,\tau}=F(\tau).
\end{equation}
The consequence (\ref{eq:48}) is
\begin{equation}
\label{eq:49}
\frac{P_{2}(\sigma)}{P_{1}(\sigma)}=const., \quad \eta=\eta(\tau).
\end{equation}
Applying (\ref{eq:3}), (\ref{eq:5}), (\ref{eq:26}), (\ref{eq:33}), (\ref{eq:34}), (\ref{eq:49}), for (\ref{eq:42}) -- (\ref{eq:46}), we find that the function $s_{,\tau}$ depends only on the variable $\tau$, and also
\begin{equation}
\label{eq:49'}
P_{1}=const.,\quad P_{2}=const., \quad P_{3}=const.
\end{equation}

Equation (\ref{eq:26}) (eq.~(\ref{eq:14})), taking into account eqs. (\ref{eq:3}), (\ref{eq:5}), (\ref{eq:34}), (\ref{eq:42}), (\ref{eq:47}), (\ref{eq:46}), (\ref{eq:49'}), relates the integration constants $P_{1}$, $P_{2}$, $P_{3}$, and can be written as
\begin{equation}
\label{eq:49''}
P_{3}=\frac{\left(P_{2}\right)^{2}\left(c_{0}\right)^{2}}{b_{0}P_{1}}.
\end{equation}
It was noted earlier that the eq. (\ref{eq:14}), and consequently the equality found (\ref{eq:49''}), is a condition for the masslessness of the null string, i.e., a condition that on all sections of the trajectory the points of the null string move at the speed of light.

According to (\ref{eq:7}), (\ref{eq:10}), outside the region where the scalar field is ``concentrated'' (i.e., outside the thin tube of the scalar field):
\begin{equation}
\label{eq:50}
f(Q)\rightarrow f_{0}, \quad f_{,Q}\rightarrow 0, \quad \lambda(\eta)f(Q)\rightarrow 1-\alpha(\eta).
\end{equation}
Then, taking into account (\ref{eq:50}), when studying the motion of the test null string, we can apply
\begin{equation}
\label{eq:51}
\left(1+\sqrt{1-\left(c_{0}f_{,Q}\right)^{2}}\right)^{1/2}=\varepsilon_{+}\approx\sqrt{2}, \quad \left(1-\sqrt{1-\left(c_{0}f_{,Q}\right)^{2}}\right)^{1/2}\approx0.
\end{equation}

For eqs. (\ref{eq:47}), (\ref{eq:46}), the range of variation of the variables $\eta$ and $Q$, depending on the sign of the derivatives of the functions $\lambda(\eta)$ and $f(Q)$, is divided into four regions:

{\bf Region I} ($f_{,Q}>0$, $\lambda_{,\eta}>0$):
\begin{equation}
\label{eq:52}
\eta\in(-\infty, 0), \quad Q\in(0,+\infty);
\end{equation}

{\bf Region II} ($f_{,Q}<0$, $\lambda_{,\eta}>0$):
\begin{equation}
\label{eq:54}
\eta\in(-\infty, 0), \quad Q\in(-\infty, 0);
\end{equation}

{\bf Region III} ($f_{,Q}>0$, $\lambda_{,\eta}<0$):
\begin{equation}
\label{eq:56}
\eta\in(0,+\infty), \quad Q\in(0,+\infty);
\end{equation}

{\bf Region IV} ($f_{,Q}<0$, $\lambda_{,\eta}<0$):
\begin{equation}
\label{eq:58}
\eta\in(0,+\infty), \quad Q\in(-\infty, 0).
\end{equation}

Integrating eqs. (\ref{eq:42}), (\ref{eq:47}), (\ref{eq:46}), for each of the regions $\bf I$ --- $\bf IV$, taking into account eqs. (\ref{eq:49}) -- (\ref{eq:51}), we find
\begin{equation}
\label{eq:65}
\lambda(\eta)=\lambda_{0j}+\alpha_{j}P_{1}\tau,
\end{equation}
\begin{equation}
\label{eq:62}
s=s_{0j}(\sigma)+\beta_{j}\frac{\left(c_{0}\right)^{2}}{b_0}\left(\frac{P_{2}}{P_{1}}\right)^{2}(\lambda(\eta))^{-1},
\end{equation}
\begin{equation}
\label{eq:63}
f_{k}^{i}(Q)=U_{k}^{i}+\varrho_{k}^{i}\frac{P_{2}}{P_{1}}\frac{\varepsilon_{+}}{\lambda(\eta)},
\end{equation}
where $\lambda_{0j}$, $s_{0j}(\sigma)$, and $U_{k}^{i}$ are ``constants'' of integration. In eqs. (\ref{eq:65}), (\ref{eq:62}), index $j$ takes the value $0$, $1$, and for $j=0$ denotes the solution of eqs. (\ref{eq:47}), (\ref{eq:46}), in the region $\bf I, II$, and for $j=1$ the solution of eqs. (\ref{eq:47}), (\ref{eq:46}), in the region $\bf III, IV$. At the same time, the constants $\alpha_{j}$, $\beta_{j}$ take the values:
\begin{equation}
\label{eq:35}
\alpha_{0}=1/a_{0},  \quad  \alpha_{1}=-1/a_{0},
\end{equation}
\begin{equation}
\label{eq:36}
\beta_{0}=-a_{0},  \quad  \beta_{1}=a_{0}.
\end{equation}
In eq. (\ref{eq:63}), index $k$ takes the values $I$ -- $IV$, and denotes the region number, index $i$ takes the value $0$, $1$, and denotes, for $i=0$, the case $Q_{,\tau}>0$, and for $i=1$, the case $Q_{,\tau}<0$. At the same time, the constants $\varrho_{k}^{i}$ have the following values:
\begin{equation}
\label{eq:64}
\begin{array}{l}
\varrho_{I}^{0}=\varrho_{II}^{1}=\varrho_{III}^{1}=\varrho_{IV}^{0}=-a_{0}c_{0}/b_{0},\\
\\
\varrho_{I}^{1}=\varrho_{II}^{0}=\varrho_{III}^{0}=\varrho_{IV}^{1}=a_{0}c_{0}/b_{0}.
\end{array}
\end{equation}

In eq. (\ref{eq:65}), the constants $\lambda_{00}$ and $\lambda_{01}$ determine the value of the parameter $\tau$ on the boundary of the regions (i.e., when $\eta=0$). And so, fixing in (\ref{eq:65})
\begin{equation}
\label{eq:66}
\lambda_{00}=\lambda_{01}=\frac{1}{f_{0}}=const,
\end{equation}
we find that when $\eta=0$ the value of the parameter $\tau=0$. Moreover:

in regions $\bf I, II$ ($\eta<0$),
\begin{equation}
\label{eq:67}
\mbox{for} \quad \eta\in(-\infty, 0), \quad \mbox{parameter values} \quad \tau\in\left(-\frac{a_{0}}{f_{0}P_{1}}; 0\right),
\end{equation}

in regions $\bf III, IV$ ($\eta>0$),
\begin{equation}
\label{eq:68}
\mbox{for} \quad \eta\in(0,+\infty), \quad \mbox{parameter values} \quad \tau\in\left(0; \frac{a_{0}}{f_{0}P_{1}}\right).
\end{equation}

Thus, for values of the parameter $\tau$ belonging to the interval
\begin{equation}
\label{eq:68'}
\tau\in\left(-\frac{a_{0}}{f_{0}P_{1}},+\frac{a_{0}}{f_{0}P_{1}}\right),
\end{equation}
the variable $\eta$ takes values in the interval
\begin{equation}
\label{eq:68''}
\eta\in\left(-\infty,+\infty\right).
\end{equation}

The explicit form of the constants $U_{k}^{i}$ in eq. (\ref{eq:63}) is determined by the boundary conditions (i.e., the value of the functions $f(Q)$ and $\lambda(\eta)$ on the boundaries $Q=0$ and $\eta=0$). Thus, for example, for the case when the surfaces $Q=0$ and $\eta=0$ are impenetrable for the test null string, the constants $U_{k}^{i}$ have the form
\begin{equation}
\label{eq:70}
U_{I}^{0}=U_{II}^{1}=U_{III}^{1}=U_{IV}^{0}=f_{0}\left(1+\frac{P_{2}}{P_{1}}\frac{a_{o}c_{0}}{b_{0}}\varepsilon_{+}\right),
\end{equation}
\begin{equation}
\label{eq:71}
U_{I}^{1}=U_{II}^{0}=U_{III}^{0}=U_{IV}^{1}=f_{0}\left(e^{-\varsigma}-\frac{P_{2}}{P_{1}}\frac{a_{o}c_{0}}{b_{0}}\varepsilon_{+}\right).
\end{equation}

For (\ref{eq:34}), (\ref{eq:42}), (\ref{eq:47}), (\ref{eq:49'}), (\ref{eq:51}), (\ref{eq:62}), eq. (\ref{eq:27}), in regions $\bf I-IV$ takes the form
\begin{equation}
\label{eq:69}
\left(s_{0j}(\sigma)\right)_{,\sigma}=0,
\end{equation}
from which it follows
\begin{equation}
\label{eq:69'}
s_{0j}(\sigma)=Y_{j},
\end{equation}
where $Y_{j}$ are constants.

Applying (\ref{eq:64}), (\ref{eq:66}), (\ref{eq:70}), (\ref{eq:71}), for eqs. (\ref{eq:63}), we find the relation between the observed variable $z$ and the parameter $\tau$ in each of the regions $\bf I-IV$:

for case $z_{,\tau}>0$ ($Q_{,\tau}>0$)
\begin{equation}
\label{eq:72}
f^{0}_{I,IV}=f_{0}+F_{+}, \quad f^{0}_{II,III}=f_{0}e^{-\varsigma}-F_{+},
\end{equation}

for case  $z_{,\tau}<0$ ($Q_{,\tau}<0$)
\begin{equation}
\label{eq:73}
f^{1}_{II,III}=f_{0}+F_{+}, \quad f^{1}_{I,IV}=f_{0}e^{-\varsigma}-F_{+},
\end{equation}
where
\begin{equation}
\label{eq:74}
F_{+}=f_{0}\frac{a_{0}c_{0}}{b_{0}}\left(1-\frac{a_{0}}{a_{0}-f_{0}P_{1}\vert\tau\vert}\right)\frac{P_{2}}{P_{1}}\varepsilon_{+},
\end{equation}
the index on the left side of eqs. (\ref{eq:72}), (\ref{eq:73}) shows the number of the region in which this solution is implemented. You can notice that for the value
\begin{equation}
\label{eq:74a}
\tau\rightarrow\pm\frac{a_{0}}{f_{0}P_{1}},
\end{equation}
the right-hand side of eqs. (\ref{eq:72}), (\ref{eq:73}), takes on unlimited values. Namely,
\begin{equation}
\label{eq:74b}
\left.\left(1-\frac{a_{0}}{a_{0}-f_{0}P_{1}\vert\tau\vert}\right)\right\vert_{\,\tau\rightarrow\pm a_{0}/f_{0}P_{1}}\rightarrow -\infty.
\end{equation}
On the other hand, according to (\ref{eq:10}), the left-hand side of eqs. (\ref{eq:72}), (\ref{eq:73}) is a bounded function that takes values in the interval $(0,\,f_{0})$. In this case, eqs. (\ref{eq:72}), (\ref{eq:73}) provide additional restrictions on the interval of change of the parameter $\tau$. Namely, they allow to reduce the interval of possible values of the parameter $\tau$ (\ref{eq:68'}).  Reducing the interval (\ref{eq:68'}), according to (\ref{eq:68''}), will lead to a restriction on the possible values of the variable $\eta$. Or in other words, it will lead to a limitation on the values of the variables $t$ and $r$ for the test null string, and as a consequence, to the formation of a region outside of which the behavior of the test null string is not defined. This region of space is called the ``interaction zone''.

The boundaries of the interaction zone for the test null string are determined by the maximum and minimum possible values of the right-hand side of the eqs. (\ref{eq:72}) and (\ref{eq:73}). Accordingly, in regions $\bf I$ and $\bf II$ the extreme left boundary of the interaction zone is realized, and in regions $\bf III$ and $\bf IV$ the extreme right boundary of the interaction zone is realized. These limits are achieved when
\begin{equation}
\label{eq:75}
\vert\tau\vert\rightarrow \frac{a_{0}b_{0}(1-e^{-\varsigma})}{f_{0}(b_{0}P_{1}(1-e^{-\varsigma})+a_{0}c_{0}P_{2}\varepsilon_{+})}.
\end{equation}

From (\ref{eq:75}) it is clear that the ``width'' of the interaction zone depends on:

$\bullet$ Values of the initial impulses of the points of the test null string (constants $P_{1}$ and $P_{2}$);

$\bullet$ Values of the constants $f_{0}$, $a_{0}$, $b_{0}$, $c_{0}$, $\varsigma$, which characterize the distribution of the scalar field and are included in the metric functions (see eqs. (\ref{eq:3}), (\ref{eq:4}), (\ref{eq:5}), (\ref{eq:7}), (\ref{eq:6}), (\ref{eq:10}));

Taking into account (\ref{eq:69'}), the constants $s_{0j}$, in eq. (\ref{eq:62}), determine the surface (by the variable $r$), on which, during movement, the test null string and the source string ``meet'' (are located on the same surface).
 Considering eqs. (\ref{eq:65}), (\ref{eq:62}), (\ref{eq:66}), (\ref{eq:69'}), on the boundary $\eta=0$, we find
\begin{equation}
\label{eq:81}
Y_{1}=Y_{0}-2a_{0}f_{0}\frac{\left(c_{0}\right)^{2}}{b_0}\left(\frac{P_{2}}{P_{1}}\right)^{2}.
\end{equation}
For (\ref{eq:66}), (\ref{eq:81}), the variables $s$ and $\eta$, which are defined by eqs. (\ref{eq:5'}), (\ref{eq:21}), (\ref{eq:65}), (\ref{eq:62}), take the form
\begin{equation}
\label{eq:82}
\eta=t-r=\lambda^{-1}\left(\frac{1}{f_{0}}-\frac{P_{1}}{a_{0}}\vert\tau\vert\right),
\end{equation}
\begin{equation}
\label{eq:83}
s=t+r=Y_{0}-\frac{\left(c_{0}\right)^{2}}{b_0}\left(\frac{P_{2}}{P_{1}}\right)^{2}\frac{(a_{0})^{2}f_{0}}{(a_{0}-f_{0}P_{1}\vert\tau\vert)}, \, \mbox{for $\eta<0$},
\end{equation}
\begin{equation}
\label{eq:85}
s=t+r=Y_{0}-a_{0}f_{0}\frac{\left(c_{0}\right)^{2}}{b_0}\left(\frac{P_{2}}{P_{1}}\right)^{2}\left(2-\frac{a_{0}}{a_{0}-f_{0}P_{1}\vert\tau\vert}\right), \, \mbox{for $\eta>0$},
\end{equation}
where the function $\lambda^{-1}\left(\tau\right)$ is determined by the explicit form of the function $\lambda(\eta)$.

\section{Graphs of motion of the test null string}\label{sec5}

It should be noted that for analyzing the motion of a test null string, the scale of change in the $\tau$ parameter in the above solution is uninformative (see ref.~\cite{ref-79}). Then, given the invariance of the action functional for the null string with respect to reparametrization on its world surface, it is convenient to move to the new parameter $\tau'$, for example, using the equality
\begin{equation}
\label{eq:84}
\tau'=\lambda^{-1}\left(\frac{1}{f_{0}}-\frac{P_{1}}{a_{0}}\vert\tau\vert\right),
\end{equation}
or, using eqs. (\ref{eq:7}), (\ref{eq:9}),
\begin{equation}
\label{eq:94''}
\vert\tau\vert=\frac{a_{0}}{f_{0}P_{1}}\exp\left(-(\xi (\vert\tau'\vert+\epsilon'/\xi))^{-2}\right).
\end{equation}

Applying (\ref{eq:84}), (\ref{eq:94''}) to the solution of (\ref{eq:34}), (\ref{eq:72}), (\ref{eq:73}), (\ref{eq:82}) -- (\ref{eq:85}), we obtain
\begin{equation}
\label{eq:86}
t-r=\tau',
\end{equation}
for $\eta<0$
\begin{equation}
\label{eq:87}
t+r=Y_{0}-a_{0}f_{0}\frac{\left(c_{0}\right)^{2}}{b_0}\left(\frac{P_{2}}{P_{1}}\right)^{2}\acute{T},
\end{equation}
for $\eta>0$
\begin{equation}
\label{eq:88}
t+r=Y_{0}-a_{0}f_{0}\frac{\left(c_{0}\right)^{2}}{b_0}\left(\frac{P_{2}}{P_{1}}\right)^{2}\left(2-\acute{T}\right),
\end{equation}
\begin{equation}
\label{eq:90}
z=q(\sigma)+Q(\tau'),
\end{equation}
where
\begin{equation}
\label{eq:90'}
\acute{T}=\left(1-\exp\left(-\left(\xi (\vert\tau'\vert+\epsilon'/\xi)\right)^{-2}\right)\right)^{-1},
\end{equation}
and the function $Q(\tau')$ is determined by one of the following equalities:

for the case $Q_{,\tau'}>0$
\begin{equation}
\label{eq:91}
Q_{I,IV}(\tau')=f^{-1}\left(f_{0}-\left(\frac{\exp\left(-(\xi (\vert\tau'\vert+\epsilon'/\xi))^{-2}\right)}{1-\exp\left(-(\xi (\vert\tau'\vert+\epsilon'/\xi))^{-2}\right)}\right)
F\right),
\end{equation}
\begin{equation}
\label{eq:91'}
Q_{II,III}(\tau')=f^{-1}\left(f_{0}e^{-\varsigma}+\left(\frac{\exp\left(-(\xi (\vert\tau'\vert+\epsilon'/\xi))^{-2}\right)}{1-\exp\left(-(\xi (\vert\tau'\vert+\epsilon'/\xi))^{-2}\right)}\right)
F\right),
\end{equation}

for the case $Q_{,\tau'}<0$
\begin{equation}
\label{eq:92}
Q_{II,III}(\tau')=f^{-1}\left(f_{0}-\left(\frac{\exp\left(-(\xi (\vert\tau'\vert+\epsilon'/\xi))^{-2}\right)}{1-\exp\left(-(\xi (\vert\tau'\vert+\epsilon'/\xi))^{-2}\right)}\right)F\right),
\end{equation}
\begin{equation}
\label{eq:92'}
Q_{I,IV}(\tau')=f^{-1}\left(f_{0}e^{-\varsigma}+\left(\frac{\exp\left(-(\xi (\vert\tau'\vert+\epsilon'/\xi))^{-2}\right)}{1-\exp\left(-(\xi (\vert\tau'\vert+\epsilon'/\xi))^{-2}\right)}\right)
F\right).
\end{equation}
In the eqs. (\ref{eq:91}) - (\ref{eq:92'}), the form of the function $f^{-1}(\tau')$ is determined by the function (\ref{eq:10}). The index on the left side of the eqs. (\ref{eq:91}) -- (\ref{eq:92'}) shows the number of the region in which this dependency is realized,
\begin{equation}
\label{eq:93}
F=\sqrt{2}f_{0}\frac{a_{0}c_{0}}{b_{0}}\frac{P_{2}}{P_{1}}.
\end{equation}
For the values of the constants given in (\ref{eq:111'}), (\ref{eq:111''}), and the values of the constants given below
\begin{equation}
\label{eq:94}
Y_{0}=160, \, P_{1}=0.1, \, P_{2}=0.85, \, a_{0}=10^{6}, \, \epsilon=\epsilon'=10^{-7},
\end{equation}
the ``left'' and ``right'' boundaries of the interaction zone, for the test null string (see eqs.~(\ref{eq:75}), (\ref{eq:84}), (\ref{eq:94''}), (\ref{eq:86})), are achieved at
\begin{equation}
\label{eq:95}
\eta\rightarrow\pm 2.86, \quad \tau'\rightarrow\pm 2.86.
\end{equation}

\begin{figure}[ht]
\center{\includegraphics[width=7.5cm]{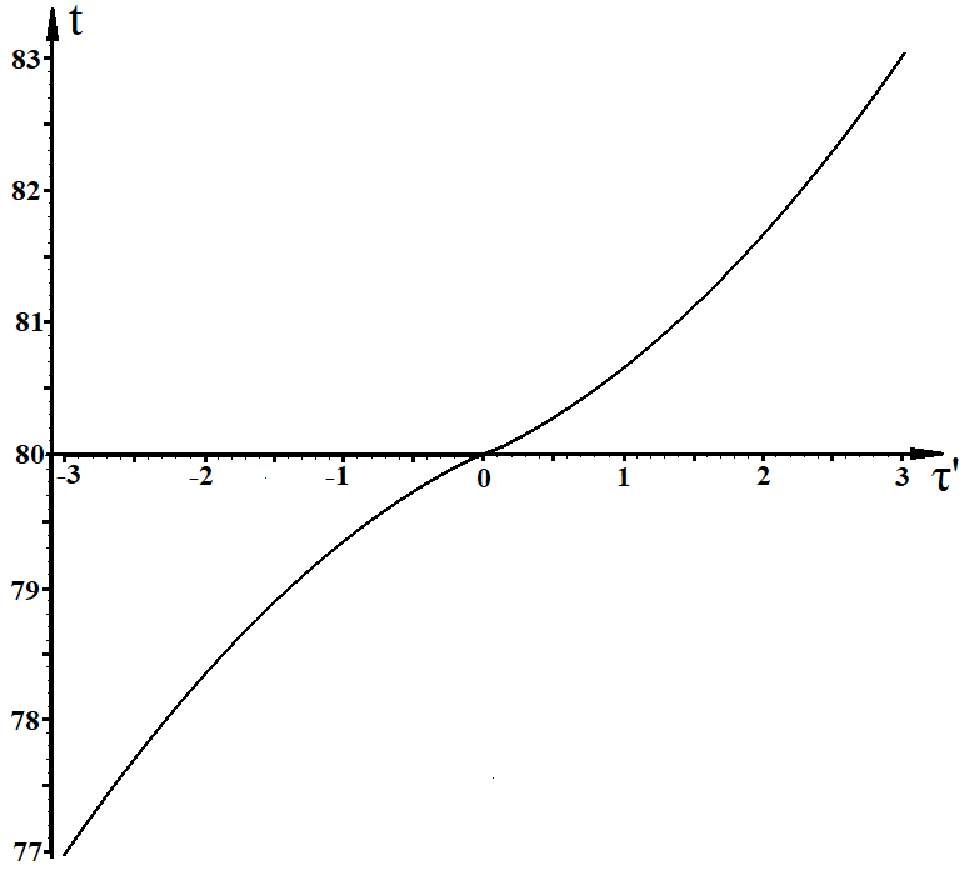}}
\hfill
\caption{\footnotesize In the figure, for the values of the constants (\ref{eq:111'}), (\ref{eq:111''}), (\ref{eq:94})), the graph of the function $t(\tau')$ is shown.}
\label{f9}
\end{figure}

\begin{figure}[ht]
\center{\includegraphics[width=7.5cm]{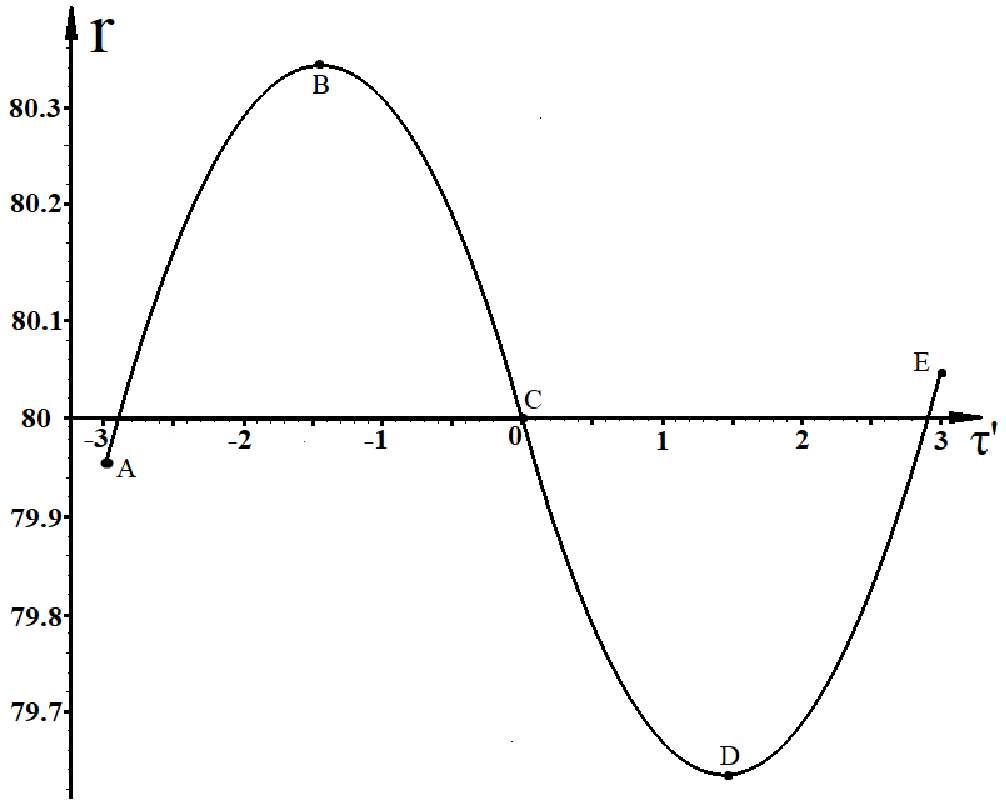}}
\hfill
\caption{\footnotesize The figure, for the values of the constants (\ref{eq:111'}), (\ref{eq:111''}), (\ref{eq:94})), shows a graph of the change in the radius of a closed test null string $r(\tau')$ as it moves in the gravitational field of the source string.}
\label{f10}
\end{figure}

\begin{figure}[ht]
\center{\includegraphics[width=9cm]{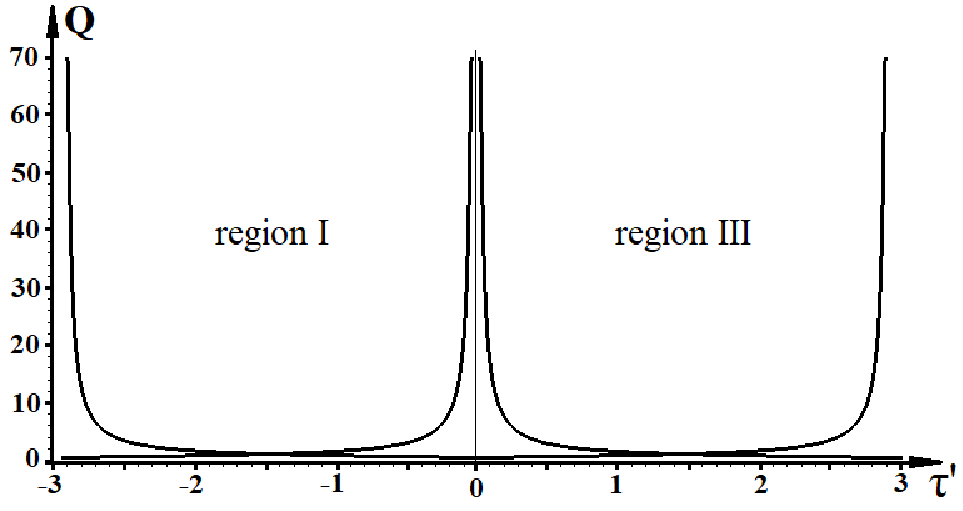}}
\hfill
\caption{\footnotesize The figure shows graphs of the dependence of the variable $Q(\tau')$, which are realized when the test null string moves in the regions $I$, $III$.}
\label{f11}
\end{figure}

\begin{figure}[ht]
\center{\includegraphics[width=9cm]{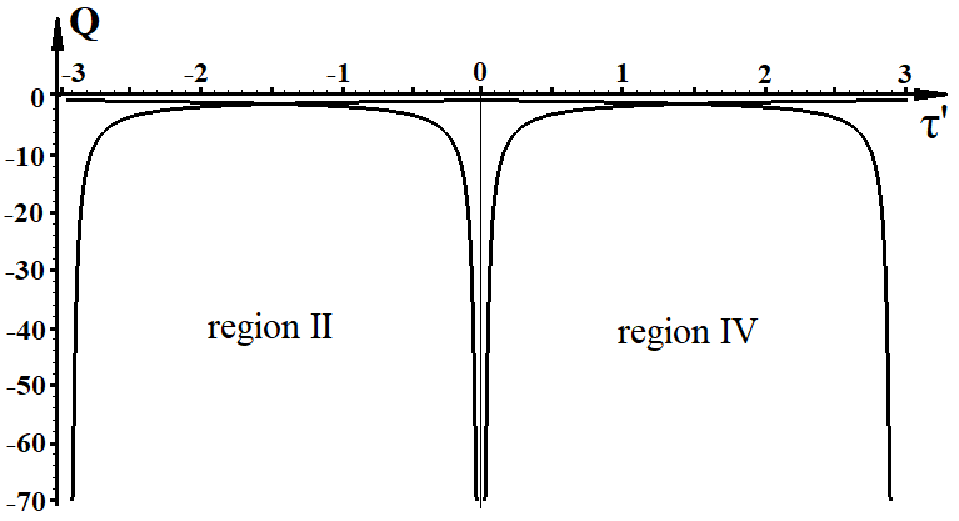}}
\hfill
\caption{\footnotesize The figure shows graphs of the function $Q(\tau')$, which are realized when the test null string moves in the regions $II$, $IV$.}
\label{f12}
\end{figure}

In Figure~\ref{f9} -- Figure~\ref{f12} for the values of the constants (\ref{eq:111'}), (\ref{eq:111''}), (\ref{eq:94})), graphs of the functions $t(\tau')$, $r(\tau')$, $Q(\tau')$ are shown, which correspond to eqs. (\ref{eq:86}) -- (\ref{eq:92'}).

The graphs of the functions $t(\tau')$, $r(\tau')$, and $Q(\tau')$ show that the gravitational field of a null string (i.e., TToMSF) can significantly influence the motion of neighboring null strings in a gas. In particular, the graph shown in Figure~\ref{f10} demonstrates that in all sections of the trajectory, the test null string experiences a ``strong'' gravitational attraction of the source string, which is capable of changing the nature of the motion of the test null string to the opposite. As confirmation of the above, Table~1 shows the values of the variables $t$ and $r$ at points $A, \ldots, E$, which are indicated in Figure~\ref{f10}, respectively, for the source string ($r_{source}$), and for the test null string ($r_{test}$).

\begin{table}[b]
\begin{center}
\begin{tabular}{cccc}
\hline
point & $t$ & $r_{source}$ & $r_{test}$ \\
\hline
$A$&$77.0$&$77.0$&$79.95$\\
\hline
$B$&$78.5$&$78.5$&$80.4$\\
\hline
$C$&$80.0$&$80.0$&$80.0$\\
\hline
$D$&$81.5$&$81.5$&$79.6$\\
\hline
$E$&$83.0$&$83.0$&$80.05$\\
\hline
\end{tabular}
\caption{\footnotesize The values of the variables $t$ and $r$ for the source string and the test null string at the points indicated in Figure~\ref{f10}.}
\end{center}
\end{table}

For the situation shown in Figure~\ref{f10}, the test null string enters the interaction zone of the source string at point $A$. At this moment in time, both the source string (see eq.~(\ref{eq:1})) and the test null string increase their size (radius), while $r_{test}>r_{source}$ (see Table~1). In the segment $A$--$C$ (see Figure~\ref{f10}) the gravitational attraction of the source string causes the test null string, which initially expanded radially (segment $A$--$B$), to change its direction of motion to the opposite at point $B$. That is, on the segment $B$--$C$ the test null string already decreases its size (radius). At point $C$ the test null string and the source string are on the same surface $\eta=t-r=0$, i.e., $r_{test}=r_{source}$ (see Table~1), but the directions of motion of these strings are different. Namely, the source string expands radially, and the test null string decreases its size (radius).

The values given in Table~1 show that on the $C$--$E$ segment $r_{test}<r_{source}$. The gravitational attraction of the source string on this segment causes the test null string at point $D$ to again change its direction of motion to the opposite. That is, on the segment $D$--$E$, the test null string, under the influence of the gravitational field of the source string, again increases its size (radius). At point $E$, the test null string leaves the source string interaction zone.

It is important to say once again that since equation (\ref{eq:14}) for the found solution (\ref{eq:87}) -- (\ref{eq:92'}) is satisfied, then for the above graphs of the functions $t(\tau')$, $r(\tau')$, and $Q(\tau')$, all points of the null string, on any time interval, move at the speed of light.

\section{Discussion}\label{sec6}

In this work we investigated a gravitational system consisting of two closed null strings. The model nature of this problem is that we fix the movement of one of the two null strings (the source string) and study the movement of the second null string (the test null string) in the gravitational field of the source string (i.e., in the field of a thin tube of a massless scalar field). In such a model, there is no way to study the gravitational influence of the test null string on the motion of the source string. But we have shown that, being in the interaction zone of the source string, the test null string experiences a ``strong'' gravitational attraction on each segment of the trajectory, which is capable of changing the nature of the motion of the test null string to the opposite. In this case, we can assume that in an ``equal'' gravitating system consisting of two closed null strings, the nature of the gravitational interaction will not change. That is, we can assume that in an ``equal'' gravitating system, each null string will experience gravitational attraction from the second null string, and this gravitational interaction can change the nature of the movement of each of these null strings to the opposite.

This feature of interaction in a gravitating system consisting of two null strings may indicate the possibility of implementing a scenario of prolonged self-consistent motion of two null strings (two closed TToMSF) within a spatially limited region. At the same time, gravitating systems consisting of two closed null strings that move self-consistently in a limited region of space can be interpreted as particles with an effective non-zero rest mass (such particles can be called ``primary particles''). Indeed, since null strings have a finite energy (see refs.~\cite{ref-274,ref-275}), then in the case of self-consistent motion of null strings this energy will be concentrated in a limited region of space, which allows us to associate with this region of space a particle with a rest mass different from zero. In this case, the gravitational interaction between individual null strings (TToMSF) can be the source of the generation of particles with an effective non-zero rest mass in the TToMSF gas. Gravitationally uniting primary particles can form macrostructures in the gas TToMSF refs.~\cite{ref-8,ref-22,ref-23}. The lifetime of primary particles should depend on their surroundings in the gas. For the same reason, the number of null strings that are simultaneously united into a macrostructure in a gas cannot be a constant value.

Earlier, in refs.~\cite{ref-270,ref-2701}, it was shown that a gas of thin tubes of a massless scalar field can be considered as one of the possible models of the dynamic scalar field of dark energy. However, the ability of such a gas, both for accelerated expansion and for the possible generation of particles with a non-zero rest mass, allows us to classify the gas of thin tubes of a massless scalar field as a model of a unified dark fluid refs.~\cite{ref-338}-\cite{ref-342}.

In our opinion, the ability of TToMSF gas to generate particles with non-zero rest mass may indicate the possibility of TToMSF gas realizing a ``strong'' coupling with baryonic matter. By ``strong'' coupling we mean the possibility of at least some part of the particles belonging to SMoPP being of a null string nature. We must immediately note that this is only an assumption, which at this point in time has no evidence. But understanding the nature of such particles we can assume experimental observations that could confirm this hypothesis. Namely, on very small time scales, the mass of particles of null-string nature should become a variable quantity depending on the time scale of observation.

The existence of its own interaction zone (see equality (\ref{eq:75}) and the explanation to it) for each thin tube of a massless scalar field in a gas can have a number of physical consequences. For example, TToMSFs that are outside each other's interaction zone in the gas will move along trajectories that locally correspond to different symmetries and thus do not gravitationally influence each other. Such local freedom in the realization of various symmetries of space-time at the micro level should indicate the possibility of realizing a homogeneous and isotropic phase in a gas of thin tubes of a massless scalar field refs.~\cite{ref-270,ref-2701}. At the same time, since the gravitational influence of an individual thin tube of a scalar field in a gas is limited only by its interaction zone, the oscillatory motion of such objects can only lead to a change in conditions at the boundary of their interaction zone, i.e., oscillations of thin tubes of a scalar field in a gas cannot lead to the generation of gravitational waves, and therefore to the loss of energy. As a consequence, the lifetime of the ``primary particles'' in the gas cannot be associated with the loss of energy due to the emission of gravitational waves.

Note also that the existence of an interaction zone for the null strings (TToMSF) forming the gas can mitigate the fine-tuning problem of $\bf T7$ given in section \ref{sec1}, since it provides a physically justified possibility to stop the ``branching'' of counterterms in the QFT.

In accordance with the equality (\ref{eq:75}), if the constants $f_{0}$, $a_{0}$, $b_{0}$, $c_{0}$, which are included in the metric functions of the quadratic form (\ref{eq:2}), are fixed, then the ``width'' of the interaction zone is determined by the initial impulses of the points of the test null string (constants $P_{1}$ and $P_{2}$). For example, if the constant $P_{2}$ is small, i.e.,
\begin{equation}
\label{eq:97}
 P_{2}\rightarrow 0,
\end{equation}
then for (\ref{eq:97}), the values of the parameter $\tau$ for the interaction zone (see eq. (\ref{eq:75})) belong to the interval
\begin{equation}
\label{eq:98}
\tau\in\left(-\frac{a_{0}}{f_{0}P_{1}},+\frac{a_{0}}{f_{0}P_{1}}\right).
\end{equation}
For (\ref{eq:98}), according to (\ref{eq:68'}), (\ref{eq:68''}), the values taken by the variable $\eta$ belong to the unlimited interval $\eta\in(-\infty,\,+\infty)$, i.e., the boundaries of the interaction zone, with respect to the variable $r$, are infinitely distant from the source string. On the other hand, with increasing values of the constant $P_{2}$ (see equality (\ref{eq:75})), the ``width'' of the interaction zone will decrease.

As follows from eqs. (\ref{eq:86})) -- (\ref{eq:93})), the motion of each thin tube of a massless scalar field in a gas depends both on the constants that characterize its gravitational field (constants $a_{0}$, $b_{0}$, $c_{0}$, $f_{0}$, $\epsilon'$, $\epsilon$, $\xi$, $\zeta$, $\varsigma$), and on the constants that characterize its motion (constants $P_{1}$, $P_{2}$). In this regard, it is interesting to find out the influence of these constants on the ability of such a gas to accelerated expansion, i.e., to find out the influence of these constants on the equation of state of a gas of thin tubes of a massless scalar field.

For a homogeneous and isotropic gas, its energy density $\rho$ and gas pressure $p$ are related to the components of the energy-momentum tensor by the relations
\begin{equation}
\label{eq:112}
\left\langle T_{0}^{0}\right\rangle=\rho, \: \left\langle T_{1}^{1}\right\rangle=\left\langle T_{2}^{2}\right\rangle=\left\langle T_{3}^{3}\right\rangle=-p.
\end{equation}

For a massless scalar field $\varphi$, the components of the energy-momentum tensor are defined by the equalities
\begin{equation}
\label{eq:113}
T_{\alpha\beta}=\varphi_{,\alpha}\varphi_{,\beta}-\frac{1}{2}g_{\alpha\beta}L,
\end{equation}
where $L=g^{ab}\varphi_{,a}\varphi_{,b}$, indices $\alpha,\,\beta,\,a,\,b$, take values $0,\,1,\,2,\,3$.

Applying (\ref{eq:2}) to (\ref{eq:113}), we find
\begin{equation}
\label{eq:114}
T_{0}^{0}=\frac{(\varphi_{,\eta})^2}{e^{2\nu}}+\frac{(\varphi_{,Q})^2}{2}\left(\frac{(q_{,\theta})^{2}}{B}+e^{-2\mu}\right),
\end{equation}
\begin{equation}
\label{eq:115}
T_{i}^{i}=T_{1}^{1}=-\frac{(\varphi_{,\eta})^2}{e^{2\nu}}+\frac{(\varphi_{,Q})^2}{2}\left(\frac{(q_{,\theta})^{2}}{B}+e^{-2\mu}\right),
\end{equation}
or using the explicit form of the distribution function of the scalar field (\ref{eq:6}), and metric functions (\ref{eq:3}) -- (\ref{eq:5})
\begin{equation}
\label{eq:116}
T_{0}^{0}=\frac{3}{2\chi(\alpha(\eta)+\lambda(\eta)f(Q))^{3}}\left\{\frac{1}{b_{0}}+\frac{1}{a_{0}}\vert\lambda_{,\eta}\vert(f_{0}-f(Q))^{2}\right\},
\end{equation}
\begin{equation}
\label{eq:117}
T_{i}^{i}=T_{1}^{1}=\frac{3}{2\chi(\alpha(\eta)+\lambda(\eta)f(Q))^{3}}\left\{\frac{1}{b_{0}}-\frac{1}{a_{0}}\vert\lambda_{,\eta}\vert(f_{0}-f(Q))^{2}\right\}.
\end{equation}
From the given eqs.~(\ref{eq:116}), (\ref{eq:117}), it follows the connection
\begin{equation}
\label{eq:118}
-T_{i}^{i}=-T_{1}^{1}=A\cdot T_{0}^{0},
\end{equation}
where
\begin{equation}
\label{eq:119}
A=A(\eta,Q)=\left(\frac{(a_{0})^{-1}\vert\lambda_{,\eta}\vert(f_{0}-f(Q))^{2}-(b_{0})^{-1}}{(a_{0})^{-1}\vert\lambda_{,\eta}\vert(f_{0}-f(Q))^{2}+(b_{0})^{-1}}\right).
\end{equation}

From the eq. (\ref{eq:119}) it is clear that the values taken by the function $A(\eta,Q)$ depend on the values of the constants $a_{0}$ and $b_{0}$, which are included in the metric functions of quadratic form (\ref{eq:2}). The value of the constant $b_{0}$ is determined by the eq. (\ref{eq:108'}), and the value of the constant $a_{0}$ must be estimated.

The constant $a_{0}$ is one of the constants that define the boundaries of the interaction zone in eq. (\ref{eq:75}). By applying (\ref{eq:66}), (\ref{eq:75}) to eq. (\ref{eq:65}) one can determine the values of the variable $\eta$ at the boundaries of the interaction zone
\begin{equation}
\label{eq:123}
\lambda(\vert\grave{\eta}\vert)=\frac{1}{f_{0}}-\frac{1}{f_{0}}\left(1+\sqrt{2}\frac{a_{0}c_{0}}{b_{0}\left(1-e^{-\varsigma}\right)}\frac{P_{2}}{P_{1}}\right)^{-1},
\end{equation}
where $\grave{\eta}$ are the values of the variable $\eta$ on the left and right boundaries of the interaction zone. We can say that for an ideal gas of thin tubes of a scalar field, the width of the interaction zone should not exceed the thickness of the scalar field tube (i.e., do not exceed core radius). According to (\ref{eq:7}), (\ref{eq:9}), outside the thin tube of the scalar field $\lambda(\eta)\rightarrow 0$. Then the ``thickness'' of a thin tube of a scalar field can be conveniently defined by the equality
\begin{equation}
\label{eq:124}
\lambda(\eta)=\frac{\kappa}{f_{0}},
\end{equation}
where $\kappa$ is a positive constant ($\kappa<1$). Combining eqs. (\ref{eq:123}) and (\ref{eq:124}), we find the condition under which the ``thickness'' of the thin tube of the scalar field and the width of the interaction zone coincide
\begin{equation}
\label{eq:125}
\sqrt{2}\frac{a_{0}c_{0}}{b_{0}\left(1-e^{-\varsigma}\right)}\frac{P_{2}}{P_{1}}=\frac{\kappa}{1-\kappa}.
\end{equation}
As already mentioned above, with increasing values of the constant $P_{2}$ the ``width'' of the interaction zone decreases. Then, the minimum possible ``width'' of the interaction zone, in the $c=1$ system of units, will correspond to the value $P_{2}=1$. In accordance with eqs. (\ref{eq:65}), (\ref{eq:67}), (\ref{eq:68}), the constant $P_{1}$ is a scaling factor that relates the observed variables $t$, $r$ (i.e., the variable $\eta$) with the parameter $\tau$ on the world surface of the test null string. In other words, the constant $P_{1}$ is a scaling constant in eq. (\ref{eq:65}), so without loss of generality we can set $P_{1}=1$.

Taking into account the above, eq. (\ref{eq:125}), for an ideal gas of thin tubes, can be written in the form
\begin{equation}
\label{eq:126}
\sqrt{2}\frac{a_{0}c_{0}}{b_{0}\left(1-e^{-\varsigma}\right)}=\frac{\kappa}{1-\kappa},
\end{equation}
or
\begin{equation}
\label{eq:127}
a_{0}=\frac{\kappa b_{0}}{\sqrt{2}c_{0}}\cdot \frac{1-e^{-\varsigma}}{1-\kappa}.
\end{equation}
Taking into account (\ref{eq:127}), the expression for the function $A(\eta,Q)$ (see eq.~(\ref{eq:119})), takes the form
\begin{equation}
\label{eq:128}
A=A(\eta,Q)=\left(\frac{\tilde{a}_{0}\vert\lambda_{,\eta}\vert(f_{0}-f(Q))^{2}-1}{\tilde{a}_{0}\vert\lambda_{,\eta}\vert(f_{0}-f(Q))^{2}+1}\right),
\end{equation}
where
\begin{equation}
\label{eq:129}
\tilde{a}_{0}=\sqrt{2}\frac{c_{0}}{\kappa}\cdot \frac{1-\kappa}{1-e^{-\varsigma}}.
\end{equation}
Note that for $\kappa=0.1$, and the values of the constants $c_{0}$, $\varsigma$, which are defined in (\ref{eq:111'}), (\ref{eq:111''})
\begin{equation}
\label{eq:130}
\tilde{a}_{0}\approx1.22.
\end{equation}

\begin{figure}[ht]
\center{\includegraphics[width=6.5cm]{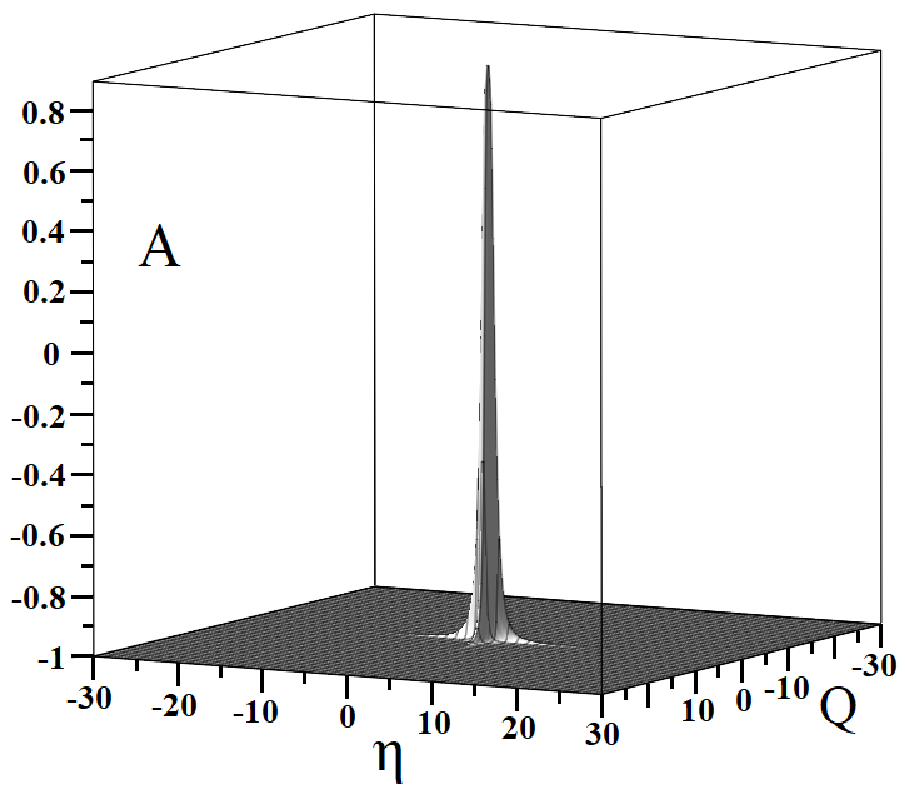}}
\hfill
\caption{\footnotesize The figure shows the graph of the function $A=A(\eta,Q)$, for the values of the constants $\zeta=\xi=\varsigma=5$.}
\label{f13}
\end{figure}

\begin{figure}[ht]
\center{\includegraphics[width=6.5cm]{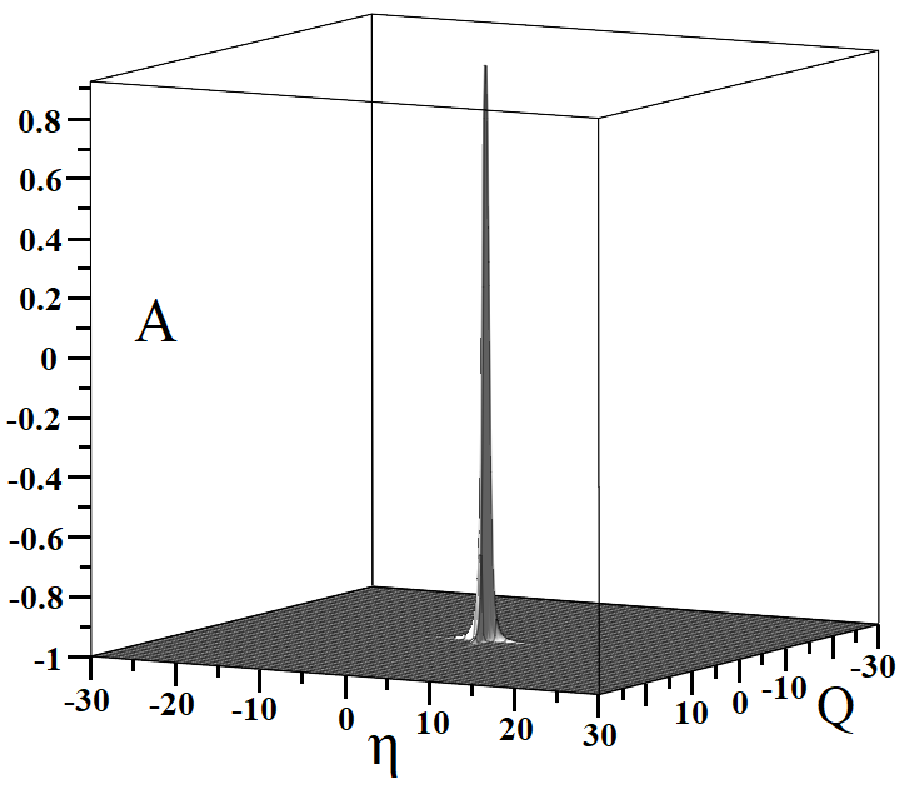}}
\hfill
\caption{\footnotesize The figure shows the graph of the function $A=A(\eta,Q)$, for the values of the constants $\zeta=\xi=\varsigma=10$.}
\label{f14}
\end{figure}

Figure~\ref{f13} and Figure~\ref{f14} show graphs (two-dimensional surfaces) of the function $A=A(\eta,Q)$, which is defined by eq. (\ref{eq:128}). In Figure~\ref{f13}, the graph of the function $A=A(\eta,Q)$ is shown for $\tilde{a}_{0}=1.22$, and the values of the constants defined in (\ref{eq:111'}), (\ref{eq:111''}), (\ref{eq:94}) (i.e., for the value of the constants $\zeta=\xi=\varsigma=5$).  In Figure~\ref{f14}, the graph of the function $A=A(\eta,Q)$ is shown for $\kappa=0.1$, $\zeta=\xi=\varsigma=10$ and, accordingly, $c_{0}=0.071$, $\tilde{a}_{0}=0.895$ (see eqs.~(\ref{eq:104}), (\ref{eq:129})).

In the graphs shown, TToMSF corresponds to the region in the vicinity of the point $\eta=Q=0$. From the graphs it is clear that everywhere outside the TToMSF region, the values of the function $A$ are constant and negative ($A=-1$). At the same time, ``inside'' the thin tube of the scalar field (i.e., in the vicinity of the point $\eta=Q=0$), the value of the function $A$ is positive $(A\rightarrow1)$. Comparing the graphs shown in Figure~\ref{f13} and Figure~\ref{f14} it is also evident that with increasing values of the constants $\zeta,\,\xi,\,\varsigma$, the region where $A\rightarrow1$ decreases, which is associated with a decrease in the thickness of TToMSF. Using the graphs of the function $A=A(\eta,Q)$ shown in Figure~\ref{f13} and Figure~\ref{f14}, for (\ref{eq:118}), we find the relation between the components of the energy-momentum tensor for a separate TToMSF in a gas. Namely, in the region that surrounds TToMSF ($\eta\neq0, Q\neq 0$),
\begin{equation}
\label{eq:130'}
-T_{i}^{i}=-T_{1}^{1}=-T_{0}^{0},
\end{equation}
at the same time, inside TToMSF ($\eta=0, Q=0$)
\begin{equation}
\label{eq:130''}
-T_{i}^{i}=-T_{1}^{1}=T_{0}^{0}.
\end{equation}

For a rarefied gas, the main contribution to the total energy-momentum tensor will come from regions of space that surround TToMSF. In addition, for a homogeneous and isotropic gas, all directions of motion are equally realizable. Then, taking into account the eq. (\ref{eq:130'}), for a rarefied, homogeneous and isotropic gas TToMSF we can write
\begin{equation}
\label{eq:120}
-\left\langle T_{1}^{1}\right\rangle=-\left\langle T_{2}^{2}\right\rangle=-\left\langle T_{3}^{3}\right\rangle \rightarrow-\left\langle T_{0}^{0}\right\rangle,
\end{equation}
or by applying eq. (\ref{eq:112}),
\begin{equation}
\label{eq:121}
p\rightarrow-\rho.
\end{equation}

For the case of a compressed gas, it can be assumed that the main contribution to the total energy-momentum tensor will be given by the regions of space located inside TToMSF, which form the gas. Then, taking into account the eq. (\ref{eq:130''}), for a supercompressed homogeneous and isotropic gas TToMSF we can write
\begin{equation}
\label{eq:122}
-\left\langle T_{1}^{1}\right\rangle=-\left\langle T_{2}^{2}\right\rangle=-\left\langle T_{3}^{3}\right\rangle \rightarrow \left\langle T_{0}^{0}\right\rangle,
\end{equation}
or by applying eq. (\ref{eq:112}),
\begin{equation}
\label{eq:122'}
p\rightarrow\rho.
\end{equation}

At the same time, as has already been said above, a decrease in the value of the initial momentum $P_{2}$ will lead to the fact that the ``width'' of the interaction zone for TToMSF in the gas will increase, which will lead to the appearance in the gas of effects associated with the peculiarity of the gravitational interaction between TToMSF that form the gas (see refs.~\cite{ref-79,ref-2701}).

\section{Acknowledgments}

We thank Pavel Kroupa for his support and attentive attitude towards our work.



\end{document}